\documentclass[aps,prl,twocolumn,superscriptaddress,floatfix]{revtex4-2}

\usepackage{graphicx}
\usepackage{amsmath,amssymb}
\usepackage[colorlinks=true,linkcolor=blue,citecolor=blue,urlcolor=blue]{hyperref}

\begin{document}
\title{Topological Enhancement of Protein Kinetic Stability}
\author{João N. C. Especial} 
\affiliation{BioISI – Instituto de Biossistemas e Ci{\^e}ncias Integrativas and Departamento de Física, Faculdade de Ci{\^e}ncias, Universidade de Lisboa, 1749-016, Lisboa, Portugal}

\author{Patrícia F. N. Faísca} 
\email{pffaisca@ciencias.ulisboa.pt}
\affiliation{BioISI – Instituto de Biossistemas e Ci{\^e}ncias Integrativas and Departamento de Física, Faculdade de Ci{\^e}ncias, Universidade de Lisboa, 1749-016, Lisboa, Portugal}

% ---------------------------------------------

\begin{abstract}
Knotted proteins embed a physical (i.e., open) knot within their native structures. For decades, significant effort has been devoted to elucidating the functional role of knots in proteins, yet no consensus has been reached. Here, using extensive Monte Carlo off-lattice simulations of a simple structure-based model, we isolate the effect of topology by comparing simulations that preserve the linear topology of the chain with simulations that allow chain crossings. This controlled framework enables us to isolate topological effects from sequence, structure and energetic contributions. We show that protein kinetic stability, defined as resistance to unfolding at a fixed temperature, is higher in knotted proteins. Additionally, kinetic stability increases significantly with knot depth, whereas foldability (or folding efficiency) is comparatively less affected. By considering a simple model of protein evolution in which amino-acid alphabet size is used as a proxy for evolutionary time, we find that increasing primary-sequence complexity through the addition of biotic amino acids predominantly enhances kinetic stability. Taken together, these results indicate that kinetic stability is a functional advantage conferred by protein knots and suggest that evolutionary pressure for kinetic stability could contribute to the persistence of knotted proteins.
\end{abstract}

\maketitle

\section{Introduction}
\label{sec:intro}
A knotted protein is a protein whose polypeptide chain forms a self-entangled configuration, embedding a physical (open) knot in its native structure. Although the first knotted protein was identified in 1977~\cite{richardson1977}, such proteins gained considerable attention only after the introduction of Taylor’s computational method for systematically determining the topological state of open polymer chains~\cite{taylor2000}. Knotted proteins account for roughly 1\% of the $\sim$225,000 entries in the Protein Data Bank (PDB)~\cite{Sulkowska2018} and are statistically rare compared with compact random loops of the same length~\cite{Grosberg2006}. The $3_1$ (trefoil) knot is by far the most prevalent, followed by the $4_1$ and $5_2$ knots~\cite{Sulkowska2018,Sulkowska2024}. Protein knots are classified as deep or shallow based on the number of residues that must be removed from a terminus to untie the chain: if more than 20 residues are required the knot is deep; otherwise it is shallow~\cite{Taylor2003}. Shallow knots are more frequent~\cite{Sulkowska2018}.

Knotted proteins are found across all three domains of life. A recent survey of 700,000 AlphaFold-predicted knotted structures across 42 proteomes reported that approximately 0.4\% of proteins in a given proteome are knotted, confirming that knots are uncommon but evolutionarily conserved~\cite{Sulkowska2024}.

Extensive theoretical and experimental work over the past two decades has provided a detailed picture of how these proteins fold (reviewed in~\cite{Faisca2015,Jackson2017,Danny2023,Tubiana2024}). Deeply knotted proteins often exhibit complex folding pathways involving parallel routes and intermediate states. Kinetically, folding is limited by a highly ordered knotting event in which the shortest terminus threads a loop formed by the remainder of the chain. This threading process generates topological barriers, backtracking, and slows down folding~\cite{Yeates2010}.

Why, then, should knotted proteins exist at all? If knotting were purely neutral or deleterious, one might expect it to be suppressed by evolution. Folding {\it in vivo} may be assisted by chaperonins such as GroEL–GroES, which increase the folding rate of deeply knotted proteins by decreasing the entropic barrier to knotting~\cite{Jackson2012,Jackson2015,faisca2016,faisca2019}. Additionally, knots may confer functional advantages to their carrier proteins. Although no consensus exists regarding their functional role~\cite{Jackson2020}, there is evidence that knots can influence active-site architecture~\cite{Nureki2002,Nureki2004,Jacobs2002}, preserve structural elements required for catalysis~\cite{Sara2024}, or regulate enzymatic activity~\cite{Ya-Ming2016}.

A major challenge in studying the role of knots is the absence of appropriate control systems~\cite{Yeates2010}. Ideally, one would compare a knotted protein with a structurally identical unknotted counterpart, but naturally occurring pairs do not exist and engineered controls inevitably introduce structural differences~\cite{Faisca2010,Soler2014,Danny2019a}. For the deeply knotted protein YibK, calorimetric measurements {\it in vitro} suggested enhanced thermal stability relative to an engineered control~\cite{Danny2019a}.
However, our previous Monte Carlo simulations addressed this problem by introducing a purely topological control:   move sets that either preserve (LTyP) or break (non-LTyP) the linear chain topology while keeping sequence, structure, and energetics identical~\cite{Especial2022,Especial2026}. Within this framework, LTyP simulations required two orders of magnitude more steps to equilibrate YibK, demonstrating that knotting strongly affects equilibration times. Yet, when equilibrium is reached, knotted and unknotted proteins exhibit the same equilibrium thermodynamic properties, including thermal stability~\cite{Especial2026}. Together with previous experimental results reporting a strong separation of timescales between unfolding (weeks) and unknotting (months)~\cite{Jennings2016}, our findings indicate that thermal denaturation experiments may yield apparent melting temperatures that do not correspond to true equilibrium distributions.\par

These results suggest that the search for the role of knots in physical properties should focus not on equilibrium thermodynamics, but on non-equilibrium properties, as well as on structural effects such as those mentioned above. An important non-equilibrium property is kinetic stability, defined as resistance to unfolding at fixed temperature.  Previous simulation studies of simple lattice~\cite{Soler2014} and off-lattice~\cite{Joanna2008} models have predicted that knots enhance kinetic stability. Here we investigate this problem by comparing simulations employing the LTyP and non-LTyP move sets. Because the protein sequence, native structure, and native-state energy are exactly the same, differences in kinetic behavior can be directly attributed to the presence of topological constraints. While the non-LTyP move set introduces nonphysical pathways absent in real proteins, it nevertheless provides an upper bound for the influence of the knot, making this approach a particularly powerful and minimally perturbative control. We find that a knot in the native structure enhances kinetic stability, and this enhancement is particularly striking in the case of deep knots. Moreover, using a simple model of protein evolution introduced in~\cite{Especial2025}, we find that kinetic stabilization provides a plausible physical basis for the evolutionary conservation of deep knots in proteins like YibK  because kinetic stability increased across evolution.

\section{Model and Methods}
\label{sec:methods}
\subsection{The C$_\alpha$ G\={o} model}
Proteins are represented by a simple C$_\alpha$ model (Figure \ref{fig:native_structure}). Accordingly, residues are reduced to hard spherical beads of uniform size, centered on the C$_\alpha$ atoms. Consecutive C$_\alpha$ atoms are connected by rigid sticks representing pseudobonds on the amide planes. We adopt a radius of 1.7~\AA{} for the beads, which is the van der Waals radius of C$_\alpha$ atoms \cite{Tsai1999}. For the length of each stick, we adopt the distance between the C$_\alpha$ atoms of the respective bonded residues in the protein's native conformation, these being approximately 2.9~\AA{} for cis bonds and 3.8~\AA{} for trans bonds.
Two non-bonded residues are said to be in contact in the native conformation if the smallest distance between any two heavy atoms, one belonging to each residue, is $\leqslant$ 4.5 \AA, this cut-off being chosen because it is slightly larger than twice the average van der Waals radius of heavy atoms in proteins.

To model protein energetics, we consider the native-centric G\={o} potential \cite{Go}. Accordingly, the total energy $E$ of a conformation defined by bead coordinates $\{\vec{r}_i\}$ is given by
%\begin{equation}
%    E\left(\{\vec{r}_i\}\right) = \varepsilon \sum_{i,j \geqslant i+2}^N
%    \left[\left(\frac{\left|\vec{r}_i - \vec{r}_j\right| - \left|\vec{r}_i^{\, nat} - \vec{r}_j^{\, nat}\right|}{w}\right)^2 +
%    1\right]^{-1}
%    \left(\chi_{ij} \chi_{ij}^{nat} + \chi_{ji} \chi_{ji}^{nat} + \frac{1}{2}\right)
%    \Delta_{ij}^{nat} \, .
%\end{equation}
%\begin{widetext}
%\begin{equation}
%E\left(\{\vec{r}_i\}\right) = \varepsilon \sum_{i,j \geqslant i+2}^N
%\left[\left(\frac{\left|\vec{r}_i - \vec{r}_j\right| - \left|\vec{r}_i^{\,\mathrm{nat}} - \vec{r}_j^{\,\mathrm{nat}}\right|}{w}\right)^2 + 1\right]^{-1}
%\left(\chi_{ij}\chi_{ij}^{\mathrm{nat}} + \chi_{ji}\chi_{ji}^{\mathrm{nat}} + \frac{1}{2}\right)
%\Delta_{ij}^{\mathrm{nat}} .
%\end{equation}
%\end{widetext}

\begin{equation}
\begin{split}
E\left(\{\vec{r}_i\}\right) = \varepsilon \sum_{i,j \geqslant i+2}^N
&\left[\left(\frac{\left|\vec{r}_i - \vec{r}_j\right|
- \left|\vec{r}_i^{\,\mathrm{nat}} - \vec{r}_j^{\,\mathrm{nat}}\right|}{w}\right)^2 + 1\right]^{-1} \\
&\times
\left(\chi_{ij}\chi_{ij}^{\mathrm{nat}} + \chi_{ji}\chi_{ji}^{\mathrm{nat}} + \frac{1}{2}\right)
\Delta_{ij}^{\mathrm{nat}} \,,
\end{split}
\end{equation}
where $N$ is the chain length measured in the number of beads, $\vec r_{i}^{nat}$ is the position vector of bead $i$ in the native structure, $\Delta_{ij}^{nat}$ is 1 if the $i-j$ contact is present in the native conformation and is 0 otherwise, $\varepsilon$ is a uniform intramolecular energy parameter (taken as $-1$ in this study, in which energies and temperatures are shown in reduced units), $w$ is the half-width of the inverse quadratic potential well, and the chirality of contact $i-j$ in the conformation under consideration is
%\begin{equation}
%    \chi_{ij} = \Theta\left( \, \left(\vec{r}_i - \vec{r}_j\right)^{ } \cdot
%                [ (\vec{r}_{j+1} - \vec{r}_j) \times
%                (\vec{r}_{j-1} - \vec{r}_j) ] \, \right) - \frac{1}{2} \, .
%\end{equation}
\begin{equation}
\chi_{ij} =
\Theta\!\left(
(\vec r_i-\vec r_j)\cdot
\big[(\vec r_{j+1}-\vec r_j)\times(\vec r_{j-1}-\vec r_j)\big]
\right)
-\frac{1}{2}\, .
\end{equation}
The chirality of the $i-j$ contact in the native conformation is
%{\small
%\begin{equation}
%\chi_{ij}^{\mathrm{nat}} =
%\Theta\!\left(
%(\vec r_i^{\,\mathrm{nat}}-\vec r_j^{\,\mathrm{nat}})\cdot
%\big[(\vec r_{j+1}^{\,\mathrm{nat}}-\vec r_j^{\,\mathrm{nat}})
%\times
%(\vec r_{j-1}^{\,\mathrm{nat}}-\vec r_j^{\,\mathrm{nat}})\big]
%\right)
%-\frac{1}{2}\, .
%\end{equation}
%}

\begin{equation}
\begin{split}
\chi_{ij}^{\mathrm{nat}} =\;
&\Theta\!\Big[
(\vec r_i^{\,\mathrm{nat}}-\vec r_j^{\,\mathrm{nat}})\cdot \\
&\qquad
\big(
(\vec r_{j+1}^{\,\mathrm{nat}}-\vec r_j^{\,\mathrm{nat}})
\times
(\vec r_{j-1}^{\,\mathrm{nat}}-\vec r_j^{\,\mathrm{nat}})
\big)
\Big]
-\frac{1}{2} .
\end{split}
\end{equation}

In equations (2) and (3), $\Theta$ is Heaviside's unit step function, which takes the value 1 if its argument is greater than zero and the value 0 otherwise.
The chirality factor in (1) favors the native conformation \emph{vis à vis} its mirror conformation. A native contact is considered formed if the distance between the centers of the respective beads differs from the distance between their C$_\alpha$ atoms in the native conformation by less than the half-width of the potential wells, $w$.

\subsection{Monte Carlo simulations}

The conformational space of the C$_\alpha$  model is explored with Monte Carlo (MC) simulations  with a move set that comprises crankshaft and pivot moves
applied to contiguous segments of the chain. At each MC step, the move type is chosen with equal probability, and the number of beads involved in the rotation is randomly selected from a uniform distribution spanning up to three quarters of the chain length. Trial conformations are accepted or rejected according to the Metropolis criterion \cite{Metropolis(1953)}. Two variants of the move set are employed. In the so-called linear-topology-preserving (LTyP) simulations, rotation amplitudes are explicitly restricted so that no bead or bond crosses another during a move, ensuring conservation of the linear topology of the chain throughout the simulation. These restrictions are implemented by computing the allowed angular interval for each move such that steric clashes and chain crossings are avoided. In non-LTyP simulations, the linear topology of the chain can be broken because rotations are unrestricted and backbone crossings are allowed; trial conformations exhibiting steric overlaps are rejected (see \cite{Especial2022} for a more details). 

Because of its low computational cosst, all equilibrium simulations are carried out with a non-LTyP move set. The melting temperature, $T_m$, is the temperature at which the heat capacity, defined in reduced units as $C_V = (<E^2> - <E>^2) / T^2$,  peaks. To determine $T_m$, which is an equilibrium parameter,  we deploy MC replica-exchange (MC-RE)~\cite{Sugita(1999)} simulations. The weighted histogram analysis method (WHAM)~\cite{Chodera(2007)} is used to analyze data from the MC-RE simulations and produce maximum likelihood estimates of the density of states from which expected values for thermodynamic properties, such as the $C_V$, are calculated as functions of temperature.   Folding progress is monitored by using the fraction of native contacts, $Q$ and the topological state (knotted or unknotted) of a sampled conformation is determined using the Koniaris-Muthukumar-Taylor (KMT) algorithm \cite{taylor2000}.

Because MC steps do not correspond to physical time, folding (and unfolding) efficiency is quantified by counting transitions rather than folding (or unfolding) rates. Specifically, we measure the number of unfolded-to-folded transitions per million MC steps (MMCs) at fixed temperature $T < T_m$, and the number of folded-to-unfolded transitions per MMCs at fixed temperature $T > T_m$, for both move sets. A conformation is considered folded if it is knotted, and its $Q$ is larger than that at which the knotting probability (defined as the fraction of knotted conformations found in an ensemble of conformations with the same  $Q$) is 0.99. 
We quantify folding efficiency by the number of unfolded-to-folded transitions at $T<T_m$, and kinetic stability as resistance to unfolding, quantified by the number of folded-to-unfolded transitions at $T>T_m$. A system that folds more efficiently exhibits a higher number of folding transitions (foldability), whereas a kinetically more stable system exhibits fewer unfolding transitions at a given temperature. The ratio of transition counts obtained from LTyP and non-LTyP simulations quantifies the extent to which knotted topology influences folding and unfolding.

\section{Results and discussion}
\label{sec:results}

\subsection{Model systems}
\label{sec:systems}
\begin{figure}[t]
    \centering
    \includegraphics[width=\linewidth]{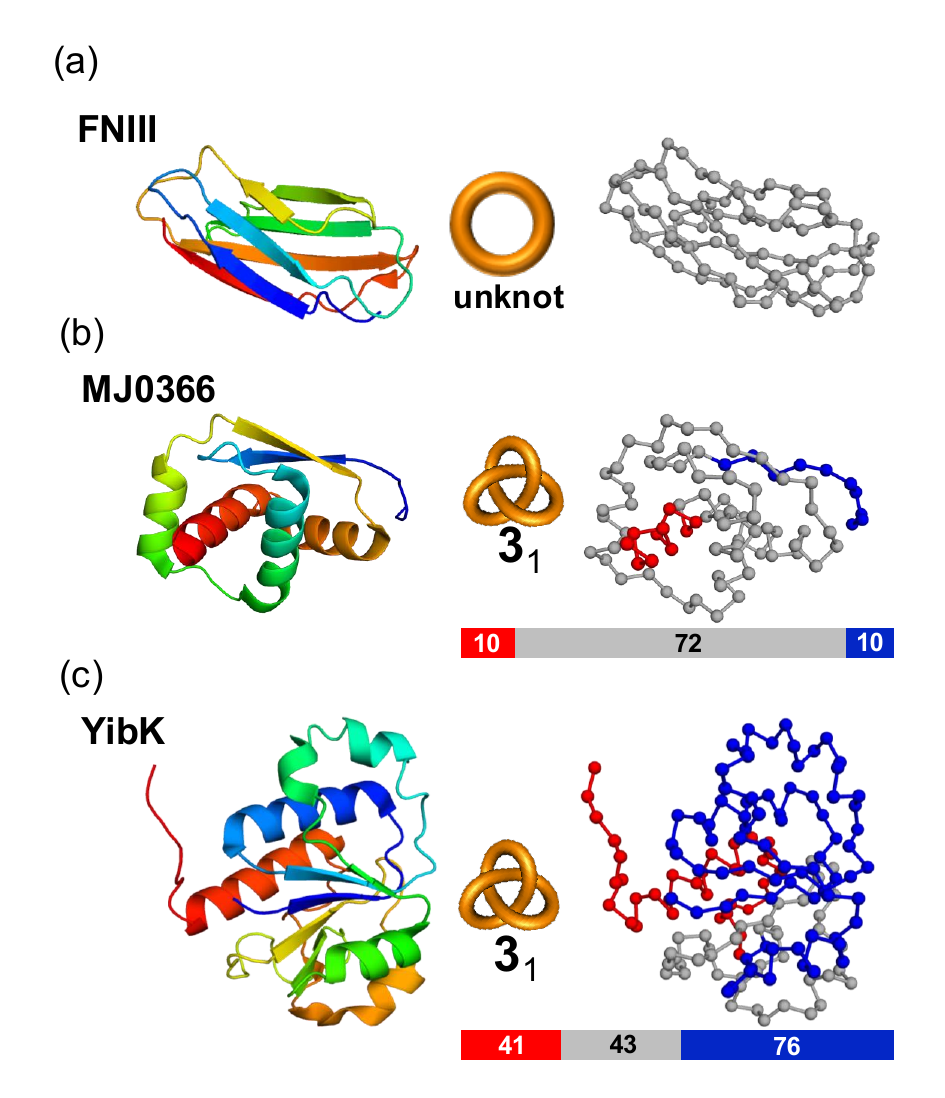}
    \caption{ Native structures of the model systems: (A) FNIII (PDB 1TEN, unknotted),  (B) MJ0366 (PDB 2EFV, shallow trefoil knot), and  (C) YibK (PDB 1J85, chain A, deep trefoil knot). 
Cartoon (left) and C$_\alpha$ bead representations (right) are shown.  The knotted core is depicted in gray; N- and C-terminal tails are colored blue and red, respectively. The sizes of the knotted core and tails (in number of beads) are indicated.
%Cartoon (left) and bead-and-stick (right) representations of the native structures of the unknotted FNIII domain  (PDB ID: 1TEN) (A), MJ0366 (PDB ID: 2EFV) (B), and YibK (PDB ID: 1J85, chain A) (C). MJ0366 contains a shallow trefoil knot, whereas YibK contains a deep trefoil knot. The knotted core is shown in gray, while knot tails are colored red (C terminus) and blue (N terminus). Each bead represents a C$_\alpha$ atom, and rigid sticks represent pseudo-bonds between pairs of C$_\alpha$ atoms. The sizes of the knotted core and knot tails (in number of beads) are indicated. 
}
 \label{fig:native_structure}
\end{figure}

We considered the $\beta$-sandwich fibronectin type III (FNIII) domain, which is unknotted (Figure \ref{fig:native_structure}A), as well as the knotted proteins MJ0366 (Figure \ref{fig:native_structure}B) and YibK (Figure \ref{fig:native_structure}C), both of which feature trefoil knots in their native structures. MJ0366 is the smallest knotted protein identified to date, with a chain length of 92 amino acids. In this protein, the knotted core, defined as the minimal polypeptide segment containing the knot, extends from amino acids 11 to 82. Because both knot tails are short (10 amino acids each), the knot is classified as shallow. In contrast, the native structure of YibK contains a deep trefoil knot. YibK is 160 amino acid long, and its knotted core spans amino acids 77 to 120, requiring the removal of 76 amino acids from the N-terminus (or 40 amino acids from the C-terminus) to untie the knot.

\begin{figure*}[t]
    \centering
    \includegraphics[width=0.8\linewidth]{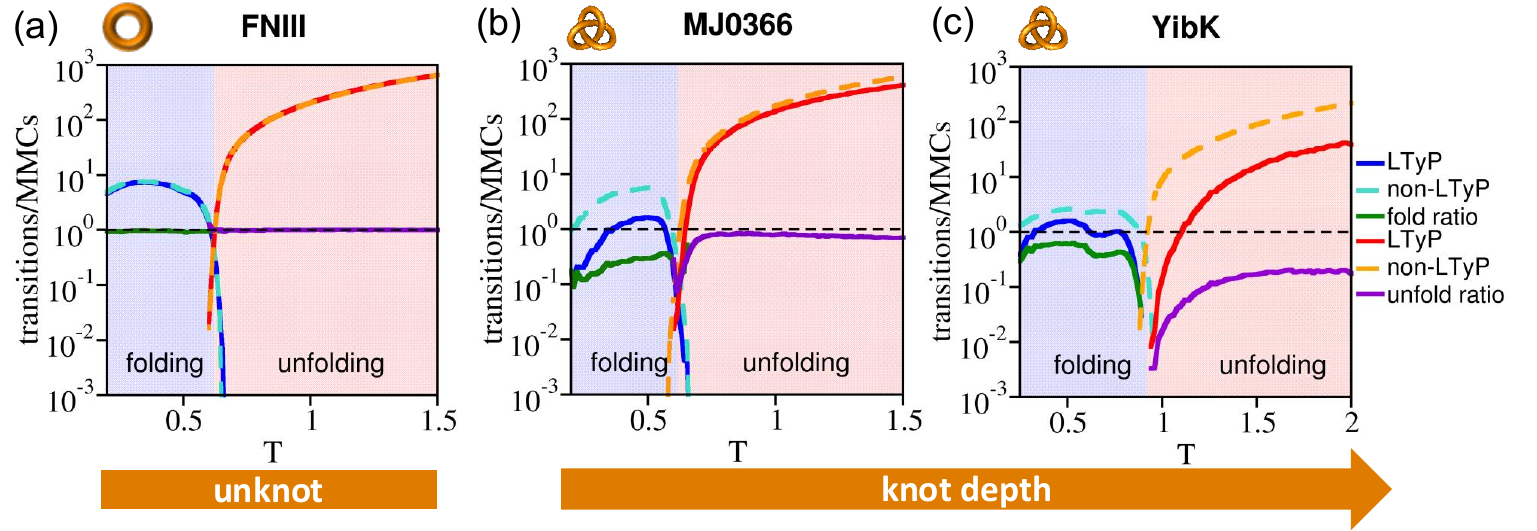}
    \caption{
Folding and unfolding transitions per million Monte Carlo steps (MMCs) as a function of temperature for 
(A) FNIII, (B) MJ0366, and (C) YibK. 
Solid lines correspond to topology-preserving (LTyP) simulations; 
dashed lines correspond to topology-breaking (non-LTyP) simulations. 
Blue and red shaded regions indicate folding ($T<T_m$) and unfolding ($T>T_m$) regimes, respectively. 
Green (fold) and purple (unfold) curves show the transition count ratios (LTyP/non-LTyP).}
   \label{fig:comparison_all}
\end{figure*}

\subsection*{Kinetic effects of knotted topology.}
Figure~\ref{fig:comparison_all} compares the number of folding and unfolding transitions per MMCs obtained from LTyP and non-LTyP simulations for the FNIII domain (Fig.~\ref{fig:comparison_all}A)  MJ0366 (Fig.~\ref{fig:comparison_all}B) and YibK (Fig.~\ref{fig:comparison_all}C). For FNIII, folding ($T<T_m$) and unfolding ($T>T_m$) transition counts are indistinguishable between the two move sets over the entire temperature range, yielding ratios close to unity. This confirms that allowing chain crossings does not affect the kinetics of unknotted proteins and, consequently, the comparison between LTyP and non-LTyP is a valid control. In contrast, knotted proteins display topology-dependent kinetics. For MJ0366, which contains a shallow trefoil knot, non-LTyP simulations yield moderately enhanced folding and unfolding transition counts compared to LTyP simulations, with ratios remaining of order unity. This indicates that, although the knot introduces kinetic constraints, their effect on foldability and kinetic stability is limited. 
A qualitatively different behavior is observed for YibK, which contains a deep trefoil knot. In this case, unfolded-to-folded transitions at $T<T_m$ are strongly suppressed under LTyP sampling, and folded-to-unfolded transitions at $T>T_m$ are reduced by more than an order of magnitude relative to non-LTyP sampling at higher temperatures. The pronounced separation between transition counts obtained from LTyP and non-LTyP simulations demonstrates that conservation of chain topology imposes a severe kinetic bottleneck and, in particular, that YibK exhibits high kinetic stability.

% \caption{Temperature dependence of folding and unfolding transitions. Number of folding and unfolding transitions per MMCs as a function of temperature ($T$) for (A) FNIII (unknotted), (B) MJ0366 (shallowly knotted), and (C) YibK (deeply trefoil knotted) proteins. Blue- and red-shaded regions indicate temperature regimes dominated by folding and unfolding transitions, respectively, with the melting temperature, $T_m$, corresponding to the interface between the two shaded regions. Solid lines correspond to LTyP simulations, while dashed lines denote non-LTyP simulations. The fold (green) and unfold (purple) ratios are defined as the ratio of transition counts obtained from non-LTyP simulations to those from LTyP simulations, thereby quantifying the extent to which knotted topology influences folding and unfolding processes.}
 \subsection*{Kinetic effects of knot depth.}
\begin{figure}[t]
    \centering
    \includegraphics[width=\linewidth]{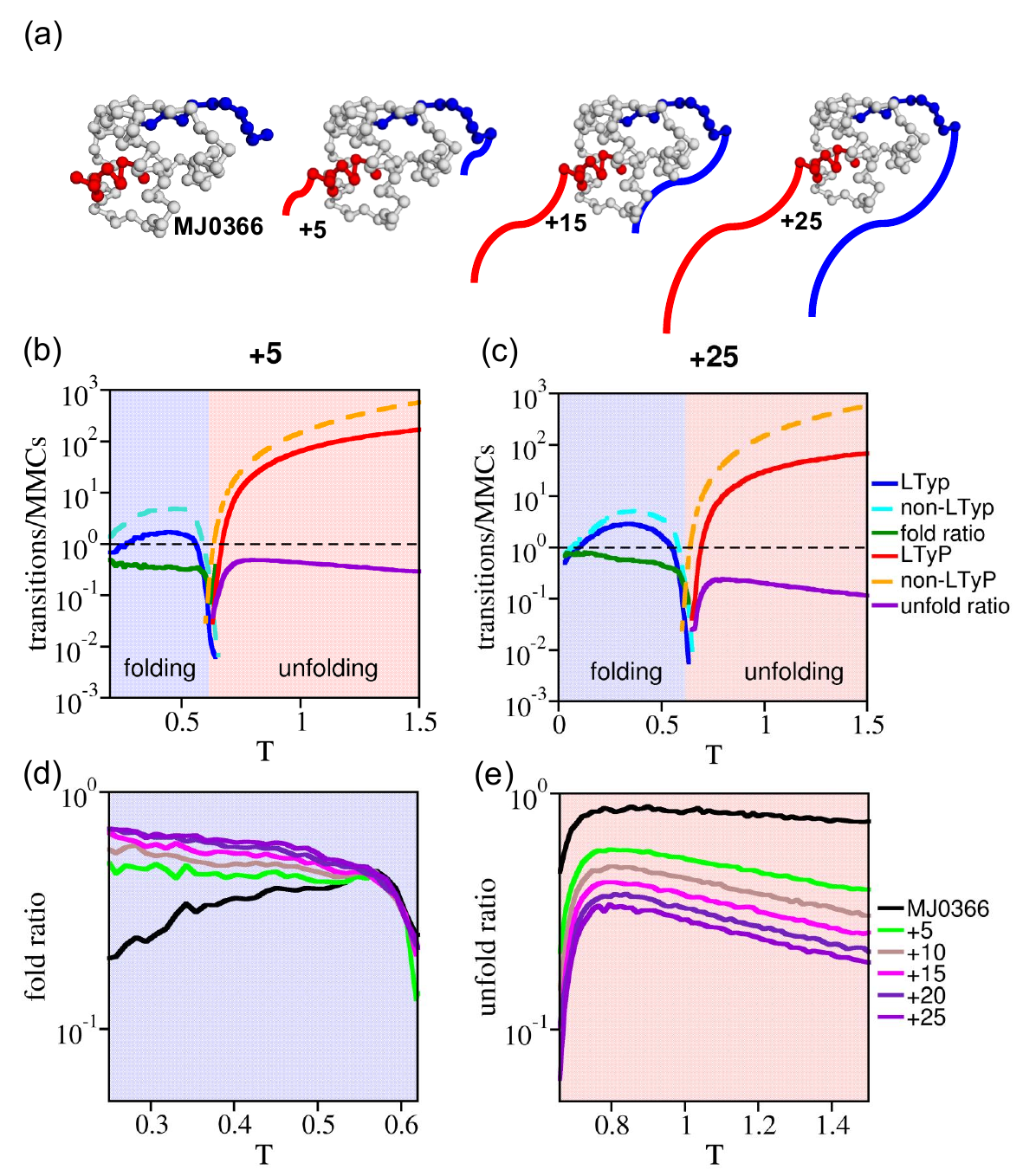}
 \caption{
Effect of increasing knot depth in MJ0366. 
(A) Native structures of MJ0366 and variants with extended terminal tails. 
(B,C) Folding ($T<T_m$) and unfolding ($T>T_m$) transitions per MMCs for MJ0366+5 and MJ0366+25 variants, respectively, obtained from LTyP (solid) and non-LTyP (dashed) simulations. 
(D) Folding transition ratios (LTyP/non-LTyP) at $T<T_m$ as a function of tail extension. 
(E) Unfolding transition ratios at $T>T_m$.
}
    \label{fig:tails}
\end{figure}

The comparison between FNIII, MJ0366, and YibK indicates that the kinetic impact of knotting increases sharply with knot depth and is negligible for shallow knots. Because these proteins also differ in chain length and native structure, we constructed MJ0366 variants with progressively extended terminal tails (5–25 beads) while preserving the native core and contact map (i.e. native state energy) (Fig.~\ref{fig:tails}A). This controlled increase in knot depth isolates its contribution to folding and unfolding kinetics. The simulations reveal a pronounced asymmetry between folding efficiency and kinetic stability as knot depth increases (Fig.~\ref{fig:tails}B,C). 

While both unfolded-to-folded transitions at $T<T_m$ and folded-to-unfolded transitions at $T>T_m$ are progressively suppressed under LTyP sampling, the effect is markedly stronger for unfolding. As the terminal tails of MJ0366 are extended, folding ratios decrease moderately and remain within the same order of magnitude, whereas unfolding ratios exhibit a much sharper and systematic reduction, reaching nearly an order-of-magnitude suppression for the longest-tail variants. This indicates that, although deeper knots reduce foldability, they impose a substantially stronger constraint on escape from the native state, thereby enhancing kinetic stability. Extended-tail MJ0366 variants thus approach the unfolding behavior of the deeply knotted YibK protein despite sharing an identical native contact map with the original MJ0366. These results demonstrate that increasing knot depth amplifies topological frustration in a highly asymmetric manner, with kinetic stability being particularly sensitive to the presence of deep knots.
 %\caption{Effect of knot depth on folding and unfolding kinetics in MJ0366. Native structures of MJ0366 and selected variants with extended termini (A). Number of folding ($T<T_m$) and unfolding ($T>T_m$) transitions per MMCs as functions of temperature for MJ0366+5 beads  (B) and MJ0366+25 beads (C), obtained from LTyP and non-LTyP simulations.  
%Folding (green) and unfolding (purple) transition ratios (non-LTyP/LTyP) (green) for the whole set of modified MJ0366 variants. Folding ratios at $T<T_m$  (D)
  %showing a monotonic suppression of folding under topology-preserving dynamics as knot depth increases (D) 
 %and unfolding ratios at $T>T_m$ (E).
 %revealing a stronger and more systematic suppression of unfolding with increasing knot depth (E).}
\subsection*{Evolution of kinetic effects of knot depth.}

\begin{figure}[t]
    \centering
    \includegraphics[width=\linewidth]{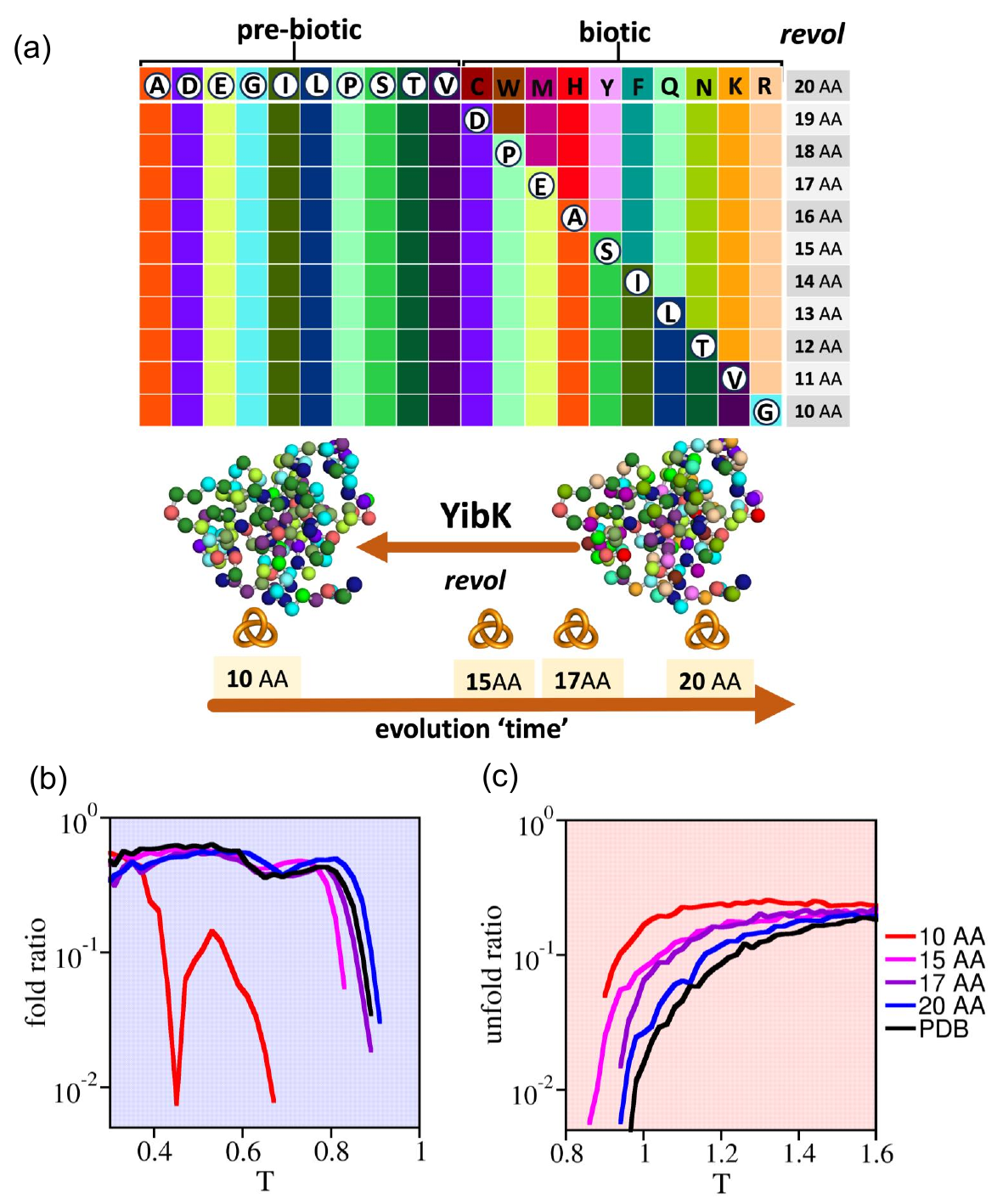}
%    \caption{Effect of amino acid alphabet size on folding and unfolding kinetics of protein YibK.
%(A) Schematic illustration of the reverse evolution ({\it revol}) procedure, in which the modern 20-letter amino acid alphabet is progressively reduced toward a putative ancestral 10-letter alphabet. Bead-and-stick representations of the native structure corresponding to the alphabet sizes 20 and 10 are shown; beads are colored according to the revol procedure.
%(B) Folding transition ratios (non-LTyP/LTyP) as functions of temperature in the folding regime ($T<T_m$) for YibK models with different alphabet sizes.
%(C) Unfolding transition ratios (non-LTyP/LTyP) as functions of temperature in the unfolding regime ($T>T_m$).}
\caption{
Effect of amino acid alphabet size on YibK kinetics. 
(A) Schematic of the reverse evolution (revol) procedure reducing the amino acid alphabet from 20 to 10 letters; representative native structures for alphabet sizes 20 and 10 are shown. 
(B) Folding transition ratios (LTyP/non-LTyP) in the folding regime ($T<T_m$). 
(C) Unfolding transition ratios in the unfolding regime ($T>T_m$).
}

    \label{fig:evol}
\end{figure}

To investigate protein folding and knotting in an evolutionary context, we adopt the reverse evolution approach introduced in Ref.~\cite{Especial2025}, in which modern protein sequences are progressively reverted toward putative ancestral sequences. In particular, alphabet size is used as a proxy for evolution ``time'', decreasing from the modern 20-letter alphabet to a reduced 10-letter alphabet representative of the prebiotic era. Reverse evolution is implemented using a nonrandom substitution scheme (termed {\it revol}) based on the relative abundances of amino acids in biotic and prebiotic conditions (Fig.~\ref{fig:evol}A), generating a series of intermediate sequences that preserve chain length while representing successive evolutionary stages. For each reverse-evolved sequence, the corresponding native structure was predicted using AlphaFold2~\cite{AlphaFold}. The predicted YibK structures closely resemble the experimentally determined PDB structure, with root mean square deviation remaining small even for the most reduced alphabets. Importantly, all predicted structures retain the same overall fold and deep trefoil knot across the entire evolutionary series.

Figure~\ref{fig:evol} compares folding (Fig.~\ref{fig:evol}B) and unfolding (Fig.~\ref{fig:evol}C) transition ratios for YibK obtained using amino acid alphabets of increasing size. In the folding regime ($T<T_m$), the reduced 10-letter alphabet exhibits a markedly lower unfolded-to-folded transition ratio under the LTyP move set. However, this reduction should be interpreted with caution: unlike the larger alphabets, the prebiotic one breaks the two-state character of the folding transition observed for the 20-letter alphabet, indicating a qualitative change in the folding landscape~\cite{Especial2025}. As a result, the transition ratio does not provide a directly comparable measure of foldability in this case. For alphabet sizes of 15, 17, and 20, where folding transitions remain two-state, increasing alphabet size systematically enhances folding efficiency, with transition ratios for the largest alphabets approaching those of the PDB-based 20-letter model.

In contrast, the unfolding regime ($T>T_m$) reveals a clearer and more systematic dependence on alphabet size. For all models, the LTyP move set suppresses folded-to-unfolded transitions, consistent with the stabilizing effect of the deep knot. Strikingly, however, the magnitude of this kinetic stabilization depends strongly on alphabet size: early, reduced-alphabet sequences exhibit substantially weaker resistance to unfolding, with transition ratios significantly closer to unity, indicating diminished kinetic stability. As the alphabet size increases, unfolding becomes progressively more suppressed, with transition ratios for the largest alphabets converging toward those of the PDB-based 20-letter model. These results indicate that the kinetic stabilization resulting from knotting is modulated by amino acid alphabet size, being significantly reduced in early, low-complexity sequences and becoming progressively more pronounced as sequence complexity increases.

\section*{Conclusions.}
By isolating topological constraints within a controlled framework, we show that the kinetic impact of knotting depends strongly on knot depth: shallow knots impose weak constraints, whereas deep knots selectively suppress unfolding and thereby enhance kinetic stability, with foldability only moderately affected. Placing these findings in an evolutionary context reveals that kinetic stability is strongly sequence-dependent: early, low-complexity sequences exhibit weaker resistance to unfolding, whereas more complex alphabets progressively enhance kinetic stabilization.

These results indicate that knots confer a functional advantage primarily through kinetic stabilization, and that this advantage becomes fully effective only when sufficient sequence specificity emerges. Such sequence-dependent stabilization provides a physical explanation for both the rarity and the evolutionary conservation of deeply knotted proteins.

%\paragraph*{Conclusions.}
%Taken together, our results show that protein knotting modulates kinetics rather than equilibrium thermodynamics. By isolating topological constraints within a controlled framework, we %demonstrate that the kinetic impact of knotting depends strongly on knot depth: shallow knots impose weak constraints, whereas deep knots markedly suppress unfolding, while folding %efficiency is less affected.

\section{Author contributions}
PFNF designed the research, prepared the figures, and wrote the manuscript. JNCE developed the code and performed all the calculations. PFNF and JNCE  analyzed the data.

\section{Declaration of interests}
The authors declare no competing interests.

\begin{acknowledgments}
Work supported by UID/04046/2025 Instituto de Biosistemas e Ciências Integrativas Centre Grant (https://doi.org/10.54499/UID/04046/2025) from FCT, Portugal. J.N.C.E. acknowledges financial support from FCT, Portugal, through PhD grant SFRH/BD/144345/2019.  A part of this work was funded by FCT through project 2022.26279.CPCA.A0. 
\end{acknowledgments}

\bibliographystyle{apsrev4-2}
\bibliography{bibliography}

\end{document}